\documentclass[numberedappendix,iop,apj]{emulateapj}
\slugcomment{{\sc \underline{Accepted for publication in Annales Geophysicae}} November 29, 2018}

\usepackage{amsmath,amssymb,amstext}
\usepackage{mhchem,ulem}
\usepackage[breaklinks,colorlinks,citecolor=blue,linkcolor=magenta]{hyperref} 

\usepackage[all]{hypcap} %Links go to figures; breaks on deluxetables (use \capstartfalse \capstarttrue to fix it)

\usepackage{natbib,color,times}
\shorttitle{Gravity Waves and Martian Ice Clouds}
\shortauthors{Y\.{I}\u{g}\.{I}t et al}
\graphicspath{{/Users/erdal/OneDrive/GWtides/Figures/}
{./Figures/}
}

\begin{document}
\title{\textbf{
Influence of gravity waves on the climatology of high-altitude Martian carbon dioxide
  ice clouds
}}

\author{Erdal Y\.{I}\u{g}\.{I}t\altaffilmark{1}, Alexander S. Medvedev\altaffilmark{2}, Paul Hartogh\altaffilmark{2}}

%\altaffiltext{1}{Department of Geophysics, Tohoku University, Sendai, Japan.}

\altaffiltext{1}{Department of Physics and Astronomy, Space Weather Laboratory,
 George Mason University, Fairfax, Virginia, USA.}

\altaffiltext{2}{Max Planck Institute for Solar System Research, 
G\"ottingen, Germany.}

%  \altaffiltext{3}{Institute of Astrophysics, Georg-August University,
% G\"ottingen, Germany.}

\begin{abstract} 
Carbon dioxide (CO$_2$) ice clouds have been routinely observed in the middle atmosphere of Mars. However, there
  are still uncertainties concerning physical mechanisms that control their altitude, geographical, and seasonal
  distributions. Using the Max Planck Institute Martian General Circulation Model (MPI-MGCM), incorporating a
  state-of-the-art whole atmosphere subgrid-scale gravity wave parameterization \citep{Yigit_etal08}, we
  demonstrate that internal gravity waves generated by lower atmospheric weather processes have wide reaching
  impact on the Martian climate. Globally, GWs cool the upper atmosphere of Mars by $\sim$10\% and facilitate
  high-altitude CO$_2$ ice cloud formation. CO$_2$ ice cloud seasonal variations in the mesosphere and the
  mesopause region appreciably coincide with the spatio-temporal variations of GW effects, providing insight into
  the observed distribution of clouds. Our results suggest that GW propagation and dissipation constitute a
  necessary physical mechanism for CO$_2$ ice cloud formation in the Martian upper atmosphere during all seasons.
\end{abstract}

% \keywords{gravity wave, thermosphere, ionosphere, sudden stratospheric warming, general circulation model, parameterizations, keyword}

\section{Introduction}  
Mars is the second most studied terrestrial planet due to its similarity and also differences to
Earth. For example, Mars is half the size of Earth, has two very exotic dwarf satellites, Phobos and Deimos; has a
similar to Earth orbital tilt, and it takes Mars nearly 1.9 Earth years to go around Sun, with a much larger
eccentricity than Earth (Appendix \ref{sec:Martian}). Thus, Mars has seasons similar to those on Earth. Studying
Mars can serve, beside the aspects of habitability, as a natural fluid dynamical laboratory, where
geophysicists can test the understanding and applicability of basic fluid dynamical principles.  Carbon dioxide
clouds have been routinely observed in the Martian atmosphere at various altitudes between $\sim$50 and $\sim$100
km \citep{ClancySandor98, Clancy_etal07, Colaprete_etal08, Maattanen_etal10, McConnochie_etal10, Vincendon_etal11,
  GonzalezGalindo_etal11, Maattanen_etal13, SeftonNash_etal13, Stevens_etal17, Aoki_etal18}.  It was hypothesized
that these so-called high-altitude clouds are formed in the regions where temperature drops below the CO$_2$
condensation threshold, which were first detected in the mesosphere during Mars Pathfinder entry and descent
\citep{Schofield_etal97}.  These high-altitude clouds are to some extent analogous to the noctilucent clouds (NLCs)
observed in Earth's mesosphere \citep{Witt62}, which are indeed high-altitude clouds. Previous numerical
simulations and observations showed that gravity wave-induced dynamical effects, such as wind fluctuations lead to
the structures observed in NLCs \citep{JensenThomas94, Rapp_etal02}.

Because the mean Martian mesosphere is in general warmer than the condensation threshold, \citet{ClancySandor98} suggested that clouds can form in pockets of cold air created occasionally by
a superposition of fluctuations associated with solar tides and gravity waves (GWs). Certainly, cold temperatures
are not the only physical mechanism required for CO$_2$ cloud formation. The microphysics calculations demonstrated
the dependence of nucleation processes on the existence and sizes of condensation nuclei \citep{Maattanen_etal10},
and that temperature excursions of several to tens of Kelvins below the condensation threshold are
required. Simulations with the Laboratoire de M\'et\'eorologie Dynamique Martian general circulation model
(LMD-MGCM) demonstrated that the spatial and temporal distributions of the predicted cold temperatures generally
correlated with the observations of high-altitude CO$_2$ clouds, but could not reproduce all of their features
\citep{GonzalezGalindo_etal11}. This study revealed the role of thermal tides in cloud formations and the authors
suggested that the discrepancies could be caused by the neglect of GWs, which were neither resolved by the MGCM,
nor accounted for in a parameterized form (with the exception of harmonics with zero horizontal phase velocities with respect to the surface generated by the flow over topography).  The
role of GWs was further addressed in the work by \citet{Spiga_etal12}, who used a mesoscale (GW-resolving) limited
area model to demonstrate for the first time with direct simulations that orographically generated waves can
propagate to the mesosphere and facilitate a creation of cold air patches at supersaturated temperatures. They
compared the distribution of a linear wave saturation index with observed clouds to find that the latter reasonably
well coincided with regions where GWs had favorable propagation conditions.

The next step in the attempt to explain the observations of high-altitude CO$_2$ clouds on the globe has been
performed with the Max Planck Institute (MPI) MGCM coupled with a whole atmosphere GW parameterization
\citep{Yigit_etal15a}.  It was shown that this technique can reproduce the occurrences of supersaturated
temperatures in low latitudes during a vernal equinox, in a good agreement with observations of mesospheric CO$_2$
clouds, which however, distinctively vary with seasons
\citep[e.g.,][]{GonzalezGalindo_etal11,SeftonNash_etal13}. In particular, the observational study of
\citet{SeftonNash_etal13} using data from NASA's Mars Climate Sounder (MCS) onboard Mars Reconnaissance Orbiter
(MRO), which demonstrated that the high-altitude clouds are continuously present in the Martian atmosphere with
distinct seasonal and latitudinal behavior. From a theoretical standpoint, it is thus instructive to study the
seasonal behavior of \ce{CO2} clouds in order to gain insight into the underlying processes.  In this paper, we
extend the approach with parameterized GWs to further assess the role of small-scale GWs in shaping spatial and
seasonal variations of pockets of cold air, which are pre-requisites for CO$_2$ cloud formation
\citep{Listowski_etal14}.

Propagation of GWs into the thermosphere has been studied extensively for Earth using idealized wave models
\citep[e.g.,][]{HickeyCole88, Walterscheid_etal13} and general circulation models (GCMs)
\citep[e.g.,][]{Yigit_etal09, Miyoshi_etal14}. They explored the fundamental processes that control propagation and
dissipation of a broad spectrum of internal waves \citep{YigitMedvedev15}.  On Mars, numerical wave models
demonstrated that GWs can propagate into the upper atmosphere and produce similar significant dynamical and thermal
forcing there \citep{Parish_etal09}.  In particular, implementation of the whole atmosphere GW scheme of
\citet{Yigit_etal08} into the Max Planck Institute Martian General Circulation Model (MPI-MGCM) revealed
substantial dynamical effects (i.e., acceleration/deceleration) in the Martian upper mesosphere and lower
thermosphere around 90--130 km, in the region of interest of this study \citep{Medvedev_etal11b}. Recently, upper
atmospheric signatures of small-scale GW waves have routinely been observed \citep{Yigit_etal15b, England_etal17}.

The structure of our paper is as follows: Next section describes the methods utilized in this research, describing
the MPI-MGCM, the whole atmosphere GW parameterization and the link between clouds and waves; section
\ref{sec:mean_fields} presents an analysis of the global annual mean fields; sections \ref{sec:seas-vari-mean} and
\ref{sec:seas-vari-grav} analyze the seasonal variations of the mean fields, gravity wave
activity, and CO$_2$ cloud formation. Section \ref{sec:dis} discusses simulation results in the context of previous
research and observations. Summary and conclusions are given in section \ref{sec:conclusion}.

\section{Methodology}\label{sec:methods}
We next describe the MGCM, outline the implemented whole atmosphere GW parameterization, how it is linked to CO$_2$
cloud formation in the model, and the setup of numerical experiments.

\subsection{Martian General Circulation Model (MGCM)}
The Max Planck Institute Martian General Circulation Model (MPI-MGCM) calculates a three-dimensional time-dependent
evolution of the horizontal and vertical winds, temperature and density of the neutral atmosphere by solving the
momentum, energy and continuity equations on a globe. The present state of the model is the result of incremental
historical development. It contains the physical parameterizations of the earlier versions
\citep{Hartogh_etal05,Hartogh_etal07,MedvedevHartogh07} and the spectral dynamical solver introduced in the work of
\citet{Medvedev_etal11a}. Of particular relevance to the subject of this paper are the parameterizations of CO$_2$
condensation/sublimation and the radiative heating/cooling scheme due to IR transfer
by CO$_2$ molecules under the breakdown of the local thermodynamic equilibrium (non-LTE). 
The former accounts for phase transitions, sedimentation of ice particles, surface ice accumulation and seasonal 
polar ice caps, thermal and mass effects. In the latter, the atomic oxygen profile of \citet{Nair_etal94} and the CO$_2$-O quenching rate coefficient 
$k_{\nu T}=3.0\cdot 10^{-12}$cm$^3$~s$^{-1}$ were used, as described in the paper of \citet{Medvedev_etal15}.

The simulations have been performed with the T21 horizontal spectral truncation, which corresponds to $64\times32 $
grid point resolution in longitude and latitude, corresponding to approximately $5.5^\circ\times 5.5^\circ$ resolution,
respectively. The current version of the model uses 67 hybrid vertical coordinates (terrain-following in the lower
atmosphere gradually and changing to pressure-based in the upper atmosphere). Its domain extends into the
thermosphere to $3.6\times 10^6$ Pa (150--160 km, depending on solar activity, temperature, etc).

\subsection{Whole Atmosphere Gravity Wave Parameterization}\label{sec:m2}
GCMs typically have resolutions insufficient for reproducing small-scale GWs. Therefore, the influence of
subgrid-scale GWs on the larger-scale atmospheric circulation has to be parameterized. The parameterizations then
estimate the effects of unresolved GWs on the resolved, large-scale flow using first principles. The vast majority
of GW schemes have been designed for terrestrial middle atmosphere GCMs \citep[see Sect. 7]{FrittsAlexander03} and,
thus, are not well suited for dissipative media such as Earth's thermosphere and Mars' middle and upper
atmosphere. We employ a GW parameterization that is specifically developed to overcome this limitation. It was
described in detail in the work of \citet{Yigit_etal08}, and the general principles of the extension of GW
parameterizations into whole atmosphere schemes have been discussed later in the work by
\citet{YigitMedvedev13}. This scheme has extensively been tested for the terrestrial environment, e.g., see the
works by \citet{YigitMedvedev16} and \citet{YigitMedvedev17} for the recent application with the Coupled Middle
Atmosphere Thermosphere-2 (CMAT2) model. The parameterization was also used within the MPI-MGCM \citep[see
e.g.,][for recent applications]{Medvedev_etal15, Medvedev_etal16} and has recently been tested in a Venusian GCM
\citep{Brecht_etal18}.

Physically-based parameterizations usually rely on certain simplifications. In the GW scheme applied here,
information about wave phases is neglected, while covariances, including the squared amplitude, are still
evaluated.  In particular, the scheme calculates the vertical evolution of the vertical flux of GW horizontal
momentum,
$\overline{\mathbf{u}^{\prime}w^{\prime}}(z) = (\overline{u^{\prime}w^{\prime}},\overline{v^{\prime}w^{\prime}})$,
taking account for the effect of dissipation on a broad spectrum of GW harmonics. In the middle and upper
atmosphere of Mars, wave damping occurs due primarily to nonlinear wave-wave interactions (breaking and/or
saturation) and molecular diffusion and thermal conduction, which are accounted for 
through the transmissivity $\tau_i$ \citep{Yigit_etal09}:
%\begin{linenomath*}
  \begin{equation}
    \label{eq:flux}
    \overline{\mathbf{u}^{\prime}w^{\prime}}_i(z) = \overline{\mathbf{u}^{\prime}w^{\prime}}_i(z_0)
    \, \frac{\rho(z_0)}{\rho(z)} \, \tau_i(z).
  \end{equation}
%\end{linenomath*}
Here overbars denote an appropriate averaging, the subscript $i$ indicates a given GW harmonic,
$\overline{\mathbf{u}^{\prime}w^{\prime}}_i(z_0)$ are the fluxes at a certain source level $z_0$, and $\rho$ is the
mass density. This formulation requires also a prescription of the characteristic horizontal scale $\lambda_h$ of
GWs for calculating $\tau_i$.  For the reasons described in our papers \citep[e.g., see the last
paragraph of Section 4 of][]{Medvedev_etal11b}, $\lambda_h=300$ km was adopted in the simulations.  Unlike in many
conventional GW schemes, no additional intermittency factors, which are often regarded as tuning factors, are used
in our scheme, because the latter is included in averaging. The parameterization is called ``spectral", because it
considers propagation of a broad spectrum of waves with different horizontal phase velocities $c_i$ (or vertical
wavelengths). The initial momentum fluxes of the phase speeds have a Gaussian distribution \citep[][Figure
2]{Medvedev_etal11b}. Note that orographically-generated GWs are represented by a single harmonic $c=0$. The scheme
takes account of interactions between GW harmonics, rather than considering them as a mere superposition of
propagating waves. Therefore, it is sometimes called ``nonlinear".  Finally, the parameterization is characterized
as a ``whole atmosphere" one to signify its physical applicability to all atmospheric layers.

The available observational constraints on GW sources in the lower atmosphere of Mars have been discussed in the
work of \citet{Medvedev_etal11a}. First, we assume a horizontally-uniform total momentum fluxes in the troposphere
with the maximum magnitude of 0.0025 m$^2$ s$^{-2}$. Recent simulations with a high-resolution MGCM
\citep{Kuroda_etal15,Kuroda_etal16} demonstrated that the sources of small-scale waves strongly vary horizontally
and with seasons and can significantly exceed this value. Thus, the current setup allows for capturing only mean GW
effects and not full details.  Second, there is a lack of detailed knowledge of GW spectra in the Martian
atmosphere. Meanwhile, there are indications of ``universality" of these spectra \citep{Ando_etal12}. Thus, we
assume the similar spectral shape of GWs in the troposphere as on Earth. Third, we consider that the mean wind at
the source level modulates the direction of propagation of GW harmonics (and their phase velocity spectrum), thus
linking the GW sources to the meteorology of the lower atmosphere \citep{Yigit_etal09, Medvedev_etal11a}. This
launch level is around $260$ Pa ($\sim 8$ km).

In the simulations to be presented, the vertical fluxes due to subgrid-scale GWs (\ref{eq:flux}) are computed in
all grid points in a time-dependent fashion for varying atmospheric conditions. These fluxes are used for
calculating GW dynamical effects, i.e., GW-induced momentum deposition (``drag") and GW thermal effects, i.e.,
heating/cooling rates \citep{YigitMedvedev09,MedvedevYigit12}, which are interactively fed into the MGCM. In the
absence of dissipation ($\tau=1$), momentum fluxes per unit volume $\rho\overline{\mathbf{u}^{\prime}w^{\prime}}$
remain constant, and GWs do not affect the large-scale wind and temperature fields, that is, the large-scale fields
that are self-consistently resolved by the MGCM. If $\tau$ falls below unity due to dissipative effects, then GWs
influence the atmospheric circulation and thermal structure. This behavior represents the process in which GWs
interact with the background flow continuously as they propagate upward. This implementation also alleviates the
limitation of the linear breaking assumption assumed by the majority of the conventional GW parameterizations. In a
realistic atmosphere GW interactions with the background atmosphere is continuous and occurs in a nonlinear
fashion. The rate of GW dissipation/breaking, which itself depends on the simulated flow, determines the momentum
and thermal forcing.
 
\subsection{Linking Gravity Waves and Ice Clouds}\label{sec:m3}
As was described above, the GW parameterization calculates covariances of wave field variables. Of particular
interest is the amplitude of temperature fluctuations $|T^\prime|=\sqrt{\overline{T^{\prime 2}}}$. Because this
scheme does not provide phase information about the subgrid-scale GW field, instantaneous values of the
parameterized (unresolved by the model) temperature disturbances $T^\prime$ are impossible to determine. However,
$|T^\prime|$ quantitatively characterizes possible maxima of fluctuations in a given point, thus allowing for
extending the probabilistic approach to CO$_2$ cloud formation.  We assume that a cloud can form, if the total
temperature $T-|T^\prime|$ drops below a certain threshold $T_s$. Then, we define the probability $P$ of this event
as
%\begin{linenomath*}
\begin{equation}
   P(z) =
    \left\{ \begin{array}{lr}
    1  & \mathrm{if} \enskip T-|T^\prime|\le T_s \\
    0  & \mathrm{otherwise}.
   \end{array} \right.
  \label{eq:P}
\end{equation}
%\end{linenomath*}
In the paper, we loosely call it the ``probability of CO$_2$ cloud formation".  In fact, cold temperature is a
necessary, but not the sufficient condition for clouds to form. Microphysics of condensation is more complex and
involves an existence and characteristics of nuclei particles. Therefore, $P$ must be treated as a certain metric
introduced for quantifying conditions favoring formation of a cloud. Because of the probabilistic nature of
$|T^\prime|$ itself, $P$ has a meaning only after a certain averaging. For example, calculating $P$ at every model
time step within a certain time interval and dividing by the number of the step yields the probability
$0 \le \bar{P} \le 1$ as a percentage of time when cloud formation was possible.

To determine $T_s$, we consider the Clausius-Clapeyron equation that relates pressure $p$ and temperature $T$ in a
system consisting of two phases, as is the case for carbon dioxide (CO$_2$) on Mars
%\begin{linenomath*}
\begin{equation}
  \label{eq:cc}
  \bigg( \frac{dp}{dT} \bigg)_{sv} = \frac{L_{sv}}{T(\nu_{v}-\nu_{s})},
\end{equation}
%\end{linenomath*}
where $L_{sv}$ is the latent heat of sublimation (the subscripts $s$ and $v$ denote the conversion from solid to
vapor phases), $\nu_{v}$, and $\nu_{s}$ are the specific volumes for vapor and solid phases, respectively. Since
$L_{sv}$ is the heat input into the system and, thus, is positive, $\nu_{v}\gg\nu_{s}$, the sublimation pressure
curve is always positive and the latent heat is temperature independent. Thus, the vapor phase of carbon dioxide
behaves like an ideal gas and the Clausius-Clapeyron equation can be integrated to obtain the expression for the
saturation temperature $T_s$
%\begin{linenomath*}
\begin{equation}
  \label{eq:Ts}
  T_s = \bigg\{\frac{1}{T_0} - \frac{R \, \ln[p(z)/p_0]}{L_{sv}}\bigg\}^{-1},
\end{equation}
%\end{linenomath*}
where $T_0 = 136.3$ K is the reference saturation temperature at $p_0=100$ Pa and $L_{sv} = 5.9\times 10^5$ J
kg$^{-1}$.  As suggested by previous experimental constraints \citep{Glandorf_etal02} a significant degree of
supersaturation is required, if microphysics of condensation is accounted for. We employ for the saturation
pressure the value $1.35\times p$ instead of $p $ in Equation (\ref{eq:Ts}). This estimate corresponds to nuclei
particles with sizes bigger than 0.5 $\mu$m and was used in previous MGCM studies \citep[e.g.,][]{Colaprete_etal08,
  Kuroda_etal13}.  The same supersaturation threshold is applied in the condensation/sublimation scheme utilized by
the MPI-MGCM for explicitly accounting for resolved CO$_2$ phase transitions. For smaller nuclei particles, which
are expected to be present in the upper atmosphere, the degree of supersaturation increases.

% FIG 1
\begin{figure*}[t]\centering
  % trim x1 y1 x2 y2
  \includegraphics[trim=0.cm 0cm 0.cm 0cm, clip,width=0.99\textwidth]{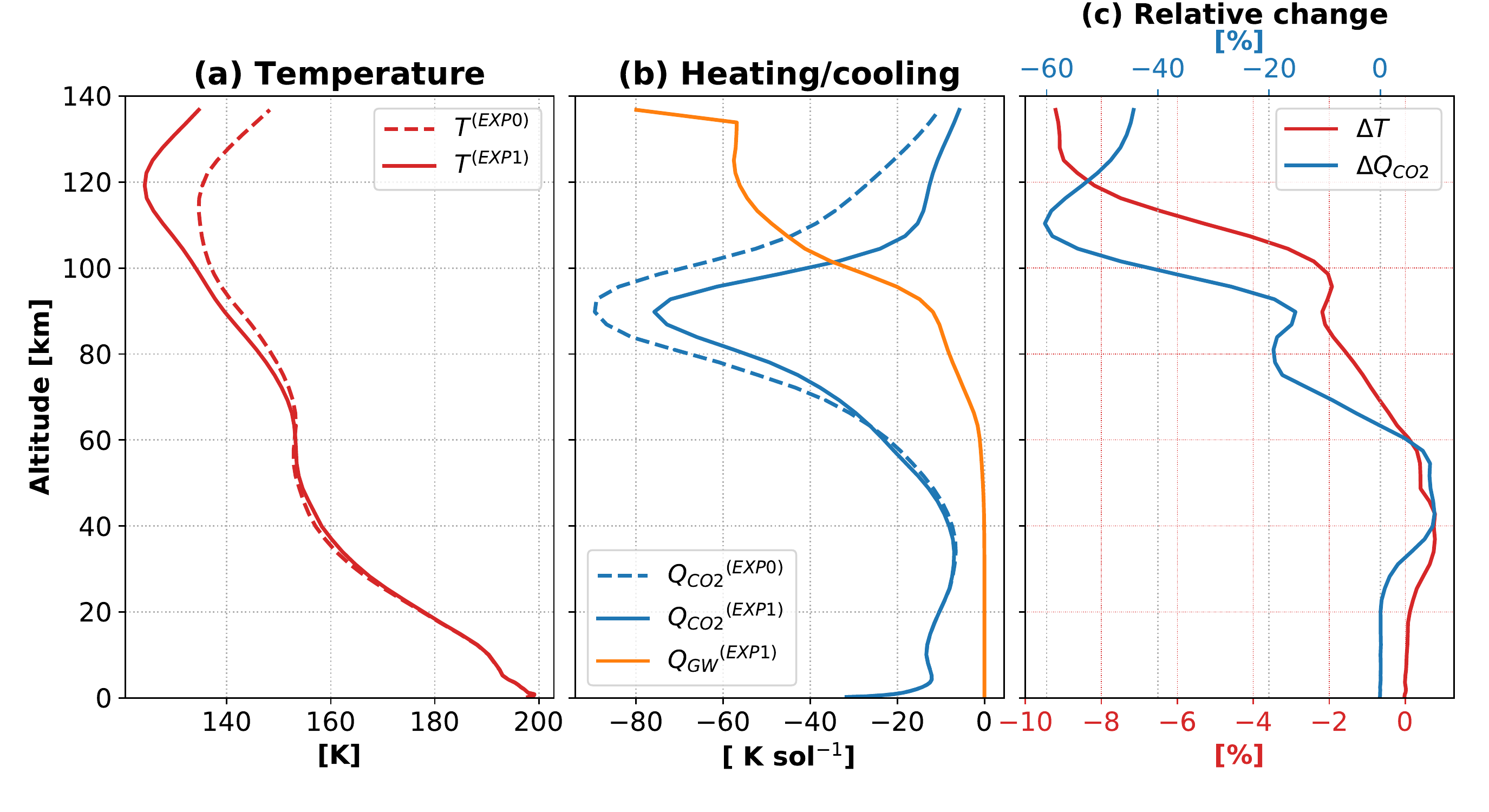}
  \caption{\textbf{Global annual mean temperature, and cooling by gravity waves and radiative
      processes in carbon dioxide molecules}. \textbf{(a)} Globally averaged annual mean
    neutral temperature ($T$ [K]) with gravity waves (solid line, ``EXP1") and without gravity
    waves (dashed lines, ``EXP0"); 
    \textbf{(b)} Globally averaged annual mean gravity wave
    heating/cooling $Q_{GW}$ [K~sol$^{-1}$] (orange line) and CO$_2$ 15
    $\mu$m cooling (blue line); 
    \textbf{(c)} Relative percentage change with respect to EXP0
    simulations for the temperature (red) and CO$_2$ cooling (blue), calculated as
    $[T^{(EXP1)}-T^{(EXP0)}]/T^{(EXP0)}$ and
    $[Q_{CO2}^{(EXP1)}-Q_{CO2}^{(EXP0)}]/Q_{CO2}^{(EXP0)}$. In both panels dashed lines
    represent the simulation without gravity waves effects, while the solid lines are for the
    simulation with gravity wave propagation from the lower atmosphere upward. The annual mean
    refers to an averaging over one Martian year (669 sols = 687 Earth days) over all
    longitudes and latitudes. }
  \label{fig:global_mean}
\end{figure*}

\subsection{Martian General Circulation Model Simulations}\label{sec:simulations}
After a multi-year spinup, the model was run for a full Martian year (669 sols $\sim$ 687 Earth days) under the
low-dust scenario and for low solar activity conditions. The dust scenario represents a composite of measurements
by the Thermal Emission Spectrometer onboard Mars Global Surveyor (MGS-TES) and the Planetary Fourier Spectrometer
onboard Mars Express (MEX-PFS) with the global dust storms removed.  Two full-Martian-year experiments have been
performed: without GWs included (EXP0) and with the GW scheme turned on (EXP1). The results to be presented are
based on daily averaged output data.
%
% FIG 2
%
\begin{figure*}[t!]\centering
  \includegraphics[trim=0.5cm 10cm 3.cm 0cm,clip,width=0.99\textwidth]{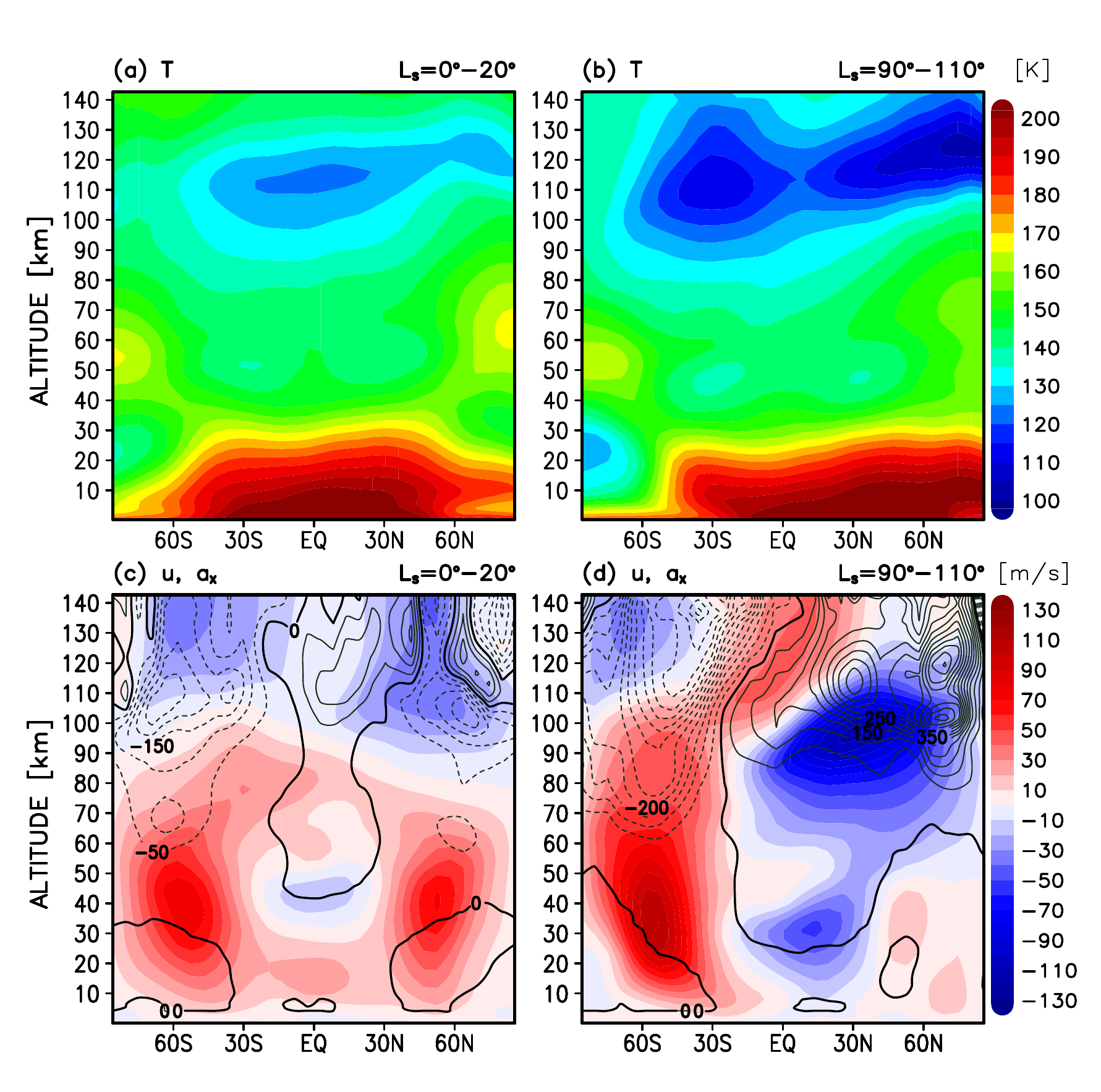}
\caption{\textbf{Altitude-latitude distributions of the mean zonal mean fields during vernal equinox
      and northern summer solstice (aphelion):} \textbf{(a)} temperature ($T$) at equinox, \textbf{(b)}
    temperature ($T$) at solstice \textbf{(c)} zonal wind $u$ (color shaded) and zonal GW drag
    (acceleration/deceleration) $a_x$ (contour lines) \textbf{(d)} zonal wind $u$ (color shaded) and
    zonal GW drag $a_x$ (contour lines). The fields are averaged over $L_s= 0^\circ-20^\circ$ (42 sols)
    for vernal equinox and over $L_s= 90^\circ-110^\circ$(44 sols) for northern hemisphere summer
    period. Temperature is in units of K, the zonal wind is in m s$^{-1}$, and the zonal GW drag is in m
    s$^{-1}$ sol$^{-1}$. Red and blue shading in the zonal wind plot represent the easterly (westward)
    and westerly (eastward) wind systems. Dashed and solid lines for the drag are for the easterly and
    westerly wave drag in intervals of 50 m s$^{-1}$ sol$^{-1}$.}
  \label{fig:mean_fields}
\end{figure*}

%
% FIG 3
%
\begin{figure*}[t!]\centering
 \vspace{-0cm}
 \includegraphics[width=0.99\textwidth]{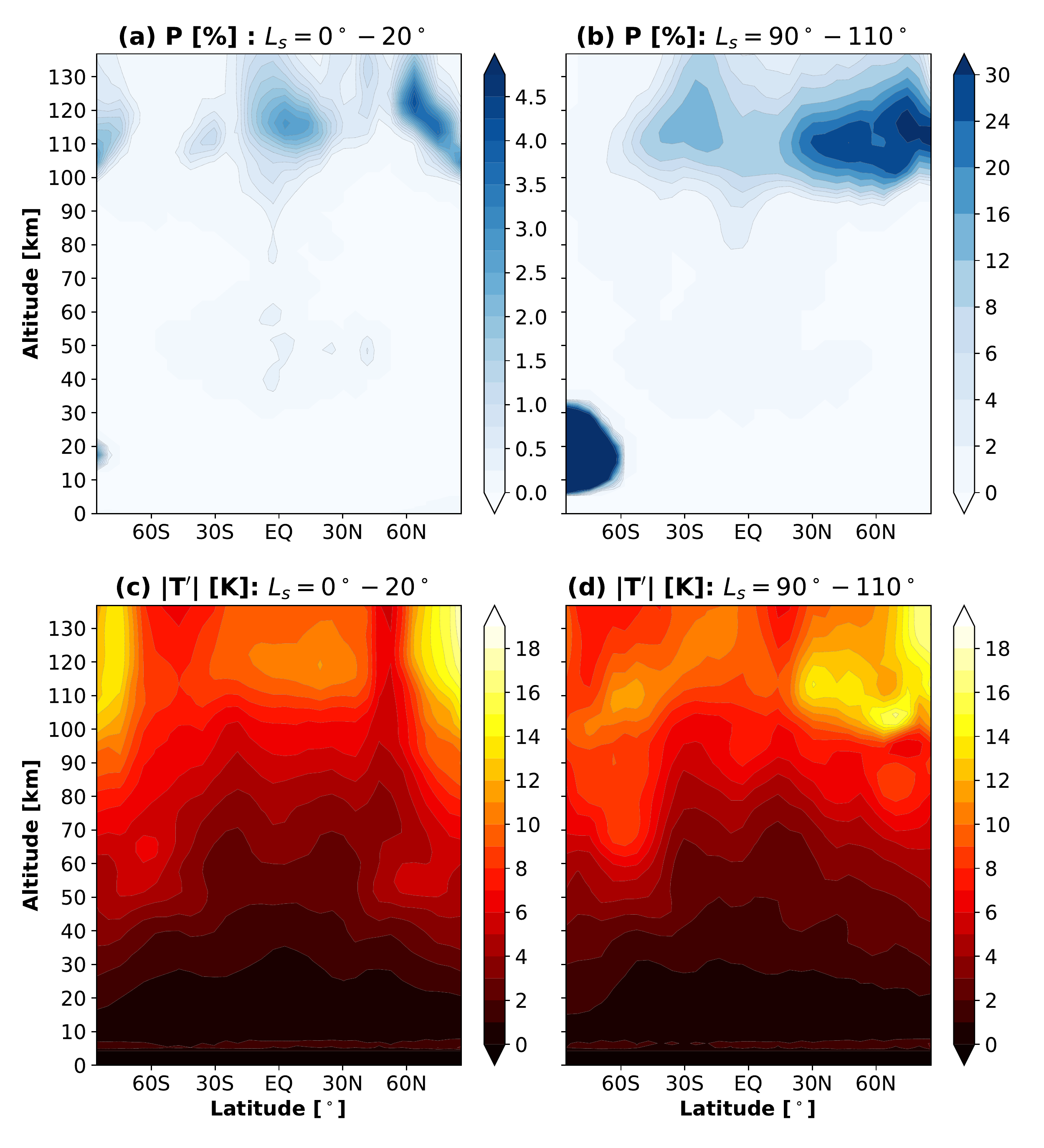}
\caption{\textbf{Altitude-latitude distributions of mean zonal mean cloud probability and
     gravity wave effects:} \textbf{(a)} cloud probability (P) at equinox, \textbf{(b)} cloud
   probability at solstice, \textbf{(c)} gravity wave induced temperature fluctuations
   ($|T^\prime| = T^\prime_{gw}$) at equinox, (d) gravity wave induced temperature fluctuations at
   solstice. The fields are averaged over a period of $L_s=0^\circ-20^\circ$ (42 sols) for
   vernal equinox and over $L_s=90^\circ-110^\circ$ (44 sols) for northern hemisphere summer
   solstice seasons, i.e., in the same manner as the data presented in \textbf{Figure}
   \ref{fig:mean_fields}. Temperature fluctuations are in units of K and the probability is
   expressed in terms of percentage.}
  \label{fig:gw_cloud_z_lat}
\end{figure*}

%
% FIG 4
% 
\begin{figure*}[t!]\centering
\vspace{-1cm}
  % trim xleft,ybottom, xright,ytop
  \includegraphics[width=0.99\textwidth]{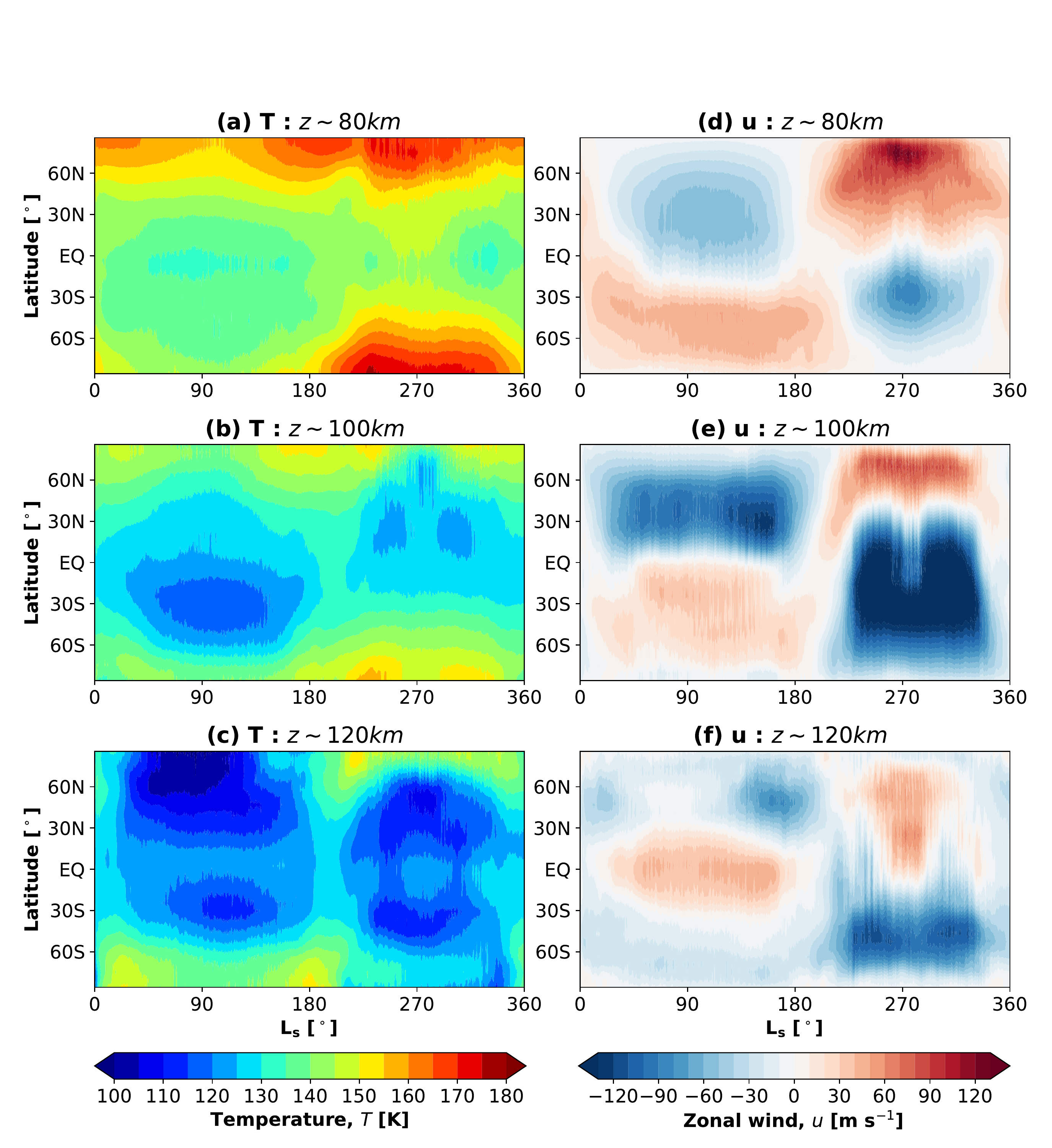}
\caption{\textbf{Seasonal variations of mean (i.e., daily and zonally averaged) atmospheric fields:}
    \textbf{(a)} Temperature ($T$) at 80 km, \textbf{(b)} $T$ at 100 km, \textbf{(c)} $T$ at 120 km, \textbf{(d)}
    Zonal wind ($u$) at 80 km, \textbf{(e)} $u$ at 100 km, \textbf{(f)} $u$ at 120 km. Temperature is in K and the
    zonal wind is in m s$^{-1}$. Red/blue shading for the wind represent eastward/westward winds. A Martian year
    has about 669 sols, which is plotted in terms of solar longitude $L_s$ (in degrees) from
      $L_s=0^\circ - 360^\circ$. $L_s=0^\circ$ marks the vernal equinox in the northern hemisphere. $L_s=90^\circ$ and
      $L_s=270^\circ$ are aphelion and perihelion seasons, respectively.}
  \label{fig:mean_fields_sol}
\end{figure*}

%
% FIG 5
%
\begin{figure*}[t!]\centering
  \includegraphics[width=1.02\textwidth]{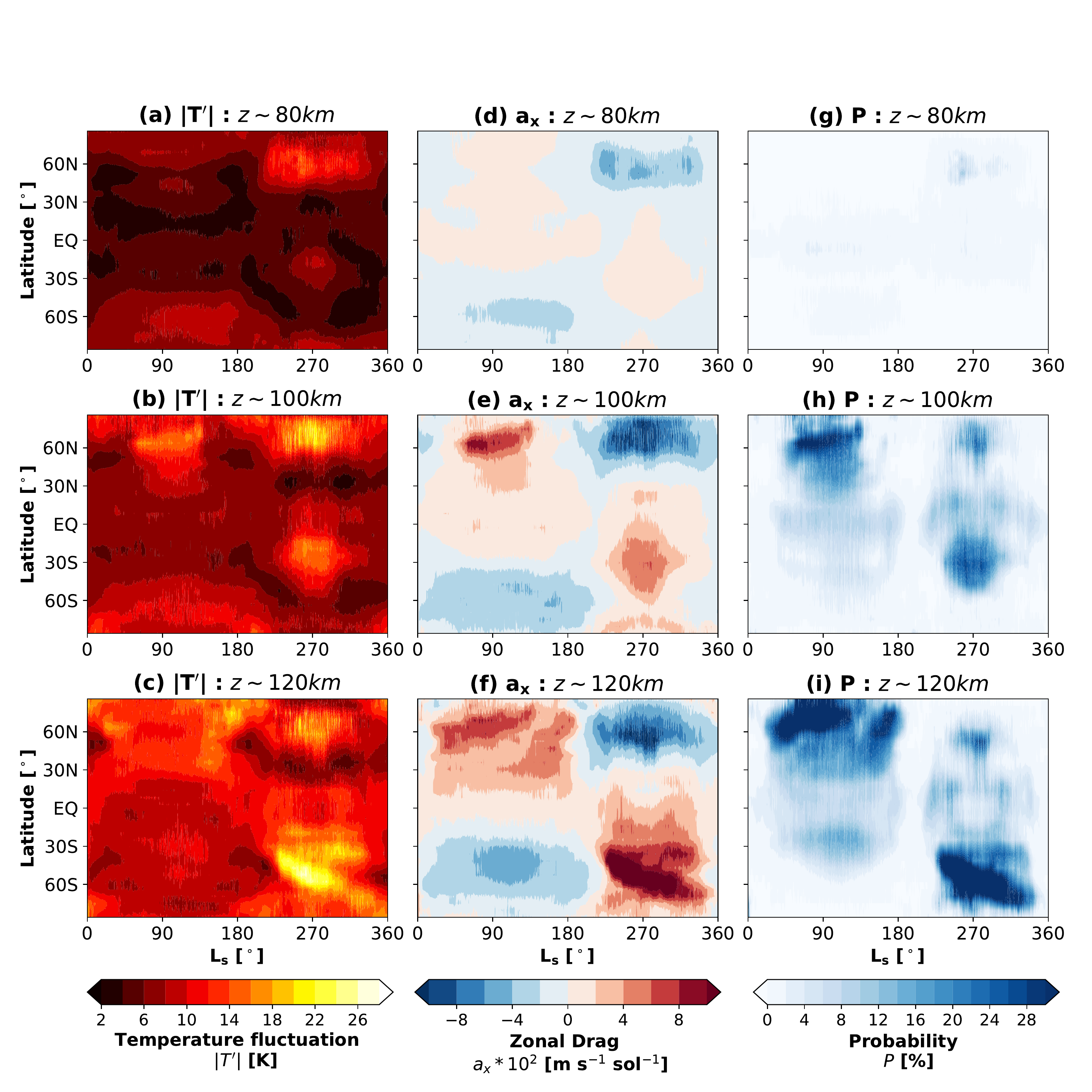}
\caption{\textbf{Seasonal variations of cloud formation probability, GW drag and GW-induced
      temperature fluctuations:} 
    \textbf{(a)} GW-induced temperature fluctuations ($|T^\prime|$) at 80 km, 
    \textbf{(b)} $|T^\prime| $ at 100 km, and
    \textbf{(c)} $|T^\prime| $ at 120 km, 
    \textbf{(d)} GW zonal drag ($a_x $) at 80 km,
    \textbf{(e)} $a_x $ at 100 km, 
    \textbf{(f)} $a_x $ at 120 km, 
    \textbf{(g)} Cloud formation probability ($P$) at 80 km,  
    \textbf{(h)} $P$ at 100 km, 
    \textbf{(i)} $P$ at 120 km.
    Probabilities are in percentage; zonal drag is in m s$^{-1}$ sol$^{-1}$, and temperature fluctuations are in K. In the drag plots (\textbf{(d-f)}, red/blue represents eastward/westward GW drag. Presented model data are in terms of daily and zonal averages. Note that the zonal drag is plotted in 200 m s$^{-1}$ sol$^{-1}$ intervals. $L_s=0^\circ$ marks the vernal equinox in the northern hemisphere.}
  \label{fig:mean_gw}
\end{figure*}

\section{Mean Fields, Gravity Waves, and Probability of CO$_2$ Ice Cloud
Formation at Solstice and Equinox}\label{sec:mean_fields}

Gravity waves can facilitate CO$_2$ cloud formation in two ways: a) by cooling down the large-scale atmosphere
globally, thus bringing its temperature closer to the condensation threshold, and b) by locally creating pockets of
cold air.  In this section, we explore the former effect by comparing the EXP0 (no-GW run) with EXP1 (GW-run)
simulations. It is instructive to compare the effects produced by GWs with the other major cooling mechanism in the
middle and upper atmosphere of Mars, -- cooling due to radiative transfer in the IR CO$_2$ bands. A detailed study
of the two mechanisms using two Martian GCMs has been performed for a vernal equinox \citep{Medvedev_etal15}. Here
our emphasis is on the global and seasonal effects.

\textbf{Figure}~\ref{fig:global_mean} presents the annual global means of the simulated temperature ($T$),
GW-induced thermal heating-cooling rates ($Q_{GW}$), and of CO$_2$ radiative cooling rates
($Q_{CO_2}$) for the experiments EXP0 (dashed line) and EXP1 (solid line). The left panel
demonstrates that inclusion of GW effects cools down the upper atmosphere at all altitudes above 60 km in a global
sense, e.g., the temperature in the mesosphere above 100 km is by $\sim$10 K lower.  Note that this change includes
both thermal and dynamical influence of GWs. The thermal one is due to GW-induced heating/cooling rates, while the
dynamical channel encompasses the temperature field response to acceleration/deceleration of the large-scale wind
by small-scale GWs. Here, it is not our goal to explore the two channels in more detail. More importantly within
the context of this paper is to demonstrate the appreciable net cooling effect of GWs.

Figure~\ref{fig:global_mean}b shows that CO$_2$ cooling is present at nearly all altitudes in the middle
atmosphere, and peaks with more than --80 K~sol$^{-1}$ around 90 km, steeply decreasing above. On the
contrary, GW cooling rates increase with altitude, exceeding that of CO$_2$ in the upper mesosphere and lower
thermosphere, and peak with --80 K~sol$^{-1}$ at 140 km. Around the mesopause and lower
thermosphere, GWs cool down the atmosphere by $\sim 5-8\%$ (Figure~\ref{fig:global_mean}c). It is also seen that
the GW-induced effects modulate the CO$_2$ cooling via changes in the background temperature: CO$_2$ cooling is
about up to 60\% weaker in the run with GWs.  In the rest of the paper, we present the results of
simulations that include GW effects (EXP1).

\textbf{Figure} \ref{fig:mean_fields} illustrates the altitude-latitude distributions of the zonal mean temperature
and wind for two characteristic seasons: the vernal equinox (averaged over 42 sols corresponding to
$L_s=0^\circ-20^\circ$, left panels) and for the aphelion solstice (44-sol average, $L_s=90^\circ-110^\circ $,
right panels). The simulated temperatures below $\sim$70--80 km are in a good agreement with observations, where
systematic satellite measurements are available \citep[e.g.,][]{Smith08}. The coldest temperatures on Mars
(favoring CO$_2$ condensation) are near the mesopause.  During the equinox, the minimum of 120 K is over the
equator. At the aphelion season, the mesopause is colder and the temperature minimum shifts to the summer
hemisphere. This behavior is tightly related to the wind distributions. It is seen that, in both seasons, zonal
jets reverse their directions near the mesopause. The similar phenomenon is well known in the mesosphere and lower
thermosphere of Earth, and is caused by the deposition of zonal momentum (i.e., zonal drag) by GWs of lower
atmospheric origin (of up to --250 m~s$^{-1}$~sol$^{-1}$ in this case), as demonstrated by the black contour
lines. During the northern hemisphere summer solstice, the asymmetry between the two hemispheres is
significant. Easterly and westerly jets dominate in the northern summer and southern winter hemispheres,
correspondingly, with the middle atmospheric jets extending higher up and reversing their directions between 110
and 120 km due to zonal GW drag acting against the mean winds. The zonal mean drag increases from $\pm 50$
m~s$^{-1}$~sol$^{-1}$ to $\pm\sim 1000 $ m~s$^{-1}$~sol$^{-1}$ from the mesosphere to the lower thermosphere, with
relatively asymmetric distribution between hemispheres. The drag of similar magnitudes has been inferred from
aerobraking data in the Martian lower atmosphere \citep{Fritts_etal06}.

\textbf{Figure} \ref{fig:gw_cloud_z_lat} demonstrates that the probability $P$ of CO$_2$ cloud formation is
strongly determined by the mean temperature. It is seen that the saturation condition for clouds are more likely
to be met during the solstice than the equinox. Specifically, the cloud formation can occur in $\sim$1\% of the
time in the equatorial mesosphere during the equinox.  Higher up at around 120 km, the cloud formation probability
increases and reaches 4.5\% with larger values found in the northern high-latitudes.  During the
solstice, the probability $P$ is larger in all atmospheric regions.  In particular, a very strong cloud formation
is seen in the winter polar lower atmosphere (Figure~\ref{fig:gw_cloud_z_lat}b), which, however, is not the focus
of this paper. In the upper atmosphere, the peak values of $P$ exceed 30\% between 100 and 120 km in the northern
hemisphere at middle- and high-latitudes.  A closer examination of Figures~\ref{fig:mean_fields} and
\ref{fig:gw_cloud_z_lat} reveals that the distributions of $P$ and mean temperature are not identical. GW-induced
fluctuations $|T^\prime|$, which are a measure of GW activity, also contribute to CO$_2$ supersaturation,
especially in low- to middle-latitudes of the middle atmosphere in both seasons and in high-altitude polar regions.
Overall, large GW-induced temperature fluctuations prevail above 100 km up to 140 km, primarily located over the
equator and in high-latitudes of both hemispheres.

\section{Seasonal Variation of the Mean Fields}
\label{sec:seas-vari-mean}

We next investigate the seasonal variations of the simulated temperature and wind in more detail by focusing on
three representative altitudes in the mesosphere and lower thermosphere: 80, 100 and 120 km.  There are too few
observations at these altitudes to date to validate the simulations. The exception is the temperature at $\sim 80$
km (\textbf{Figure}~\ref{fig:mean_fields_sol}a), which can be directly compared to retrievals from Mars Climate
Sounder (MCS) onboard Mars Reconnaissance Orbiter (MRO) \citep[see Figure 10 in the paper
of][]{SeftonNash_etal13}. Both observations and simulations demonstrate a relatively symmetric with respect to the
equator distributions during equinoxes ($L_s=0^\circ, L_s=180^\circ$).  The lowest and highest temperatures occur
in the southern hemisphere during winters and summers, correspondingly. The model generally reproduces the observed
temperature well, except that it overestimates it in the southern hemisphere winter by up to 20 K.  Alternating
with seasons zonal winds at $\sim$80 km represent an extension of the lower- and middle atmosphere jets formed as a
consequence of the Coriolis force acting on the summer-to-winter meridional circulation cell.

In the upper mesosphere (100 km, Figure~\ref{fig:mean_fields_sol}b,e) the simulated temperature and wind show
variations similar to that at 80 km, but with noticeably colder temperatures. Around the mesopause (120 km,
  Figure \ref{fig:mean_fields_sol}c,f), the simulated seasonal variations differ significantly from those in the
mesosphere. It is seen that the coldest temperatures of down to 90--100 K are found around the summer
high-latitudes at solstices, and the temperature distributions are
hemispherically less symmetric. The summer high-latitude hemispheres are remarkably different. Polar temperatures
fall down to 90 K in the summer hemisphere at the aphelion and to 115 K at the perihelion. The zonal winds reverse
their directions at 120 km, which is especially well seen during the aphelion season.  During other seasons, the
simulated winds demonstrate a significant weakening as compared to distributions in the mesosphere. The latter is
primarily attributed to the GW drag, which we present next along with GW-induced temperature fluctuations.

\section{Seasonal Variation of Gravity Wave Activity and Probability of CO$_2$ Ice Clouds}
\label{sec:seas-vari-grav}

Parameterized GW-induced temperature fluctuations ($|T^\prime|$), GW drag ($a_x$), and probability of
cloud formation ($P$) are studied next in Figure \ref{fig:mean_gw} in the same manner as temperature and zonal
winds are presented in Figure \ref{fig:mean_fields_sol}. Overall, the seasonal variations of the parameterized
GW-induced temperature fluctuations, which are created by GW harmonics that survived propagation from the lower
atmosphere depend on the assumed wave sources and on filtering by the underlying mean winds. In
the mesosphere (80 km, Figure~\ref{fig:mean_gw}a), the fluctuations of up to 16 K enhance at middle- to
high-latitudes and during the solstices with slightly larger magnitudes during winters. The middle
column of Figure~\ref{fig:mean_gw} shows the seasonal variations of the zonal GW drag, which is
largely determined by the background winds below presented in Figure~\ref{fig:mean_fields_sol} and characterizes
the rate of change of GW momentum fluxes with height. It is seen that it is directed mainly against the mean flow
throughout the mesosphere. Finally, the probability $P$ of cloud formation is plotted in the rightmost column of Figure~\ref{fig:mean_gw}. A continuous presence of $P$ of up to 2-4\% is seen
around the equator at 80 km nearly throughout the entire Martian year. After the northern summer hemisphere
solstice (aphelion), regions of cloud formation gradually expand to lower-latitudes ($\pm 30^\circ$), resembling a
fork-like structure, in some level of agreement with \citet{SeftonNash_etal13}'s observations.  During southern
winter solstice, the probability of cold pocket formation is somewhat present around midlatitudes. There is
some degree of correlation between the cloud formation probability and GW activity represented as fluctuations and
drag.

In the upper mesosphere (100 km, middle row), GW-induced temperature fluctuations increase, along with the GW
drag imposed on the mean circulation, and the cloud formation probability demonstrates a more definitive
correlation with the GW activity during all seasons. Cold pockets occur more frequently at middle- and
high-latitudes ($P \sim 16-20 \% $), exceeding the equatorial cloud probability rate. Around the mesopause,
GW-induced fluctuations increase further maximizing at middle- and high-latitudes with values of up to 26 K during
both aphelion and perihelion. The probability $P$ increases to more than $\sim$30\%
correspondingly.

\section{Discussion}\label{sec:dis}

As mentioned in the description of the model, the CO$_2$ condensation/sublimation scheme employed in the
MPI-MGCM is able to resolve CO$_2$ ice formation and annihilation when temperature in a grid point
crosses the condensation threshold. In our simulations, there were very few occurences of such clouds in
the mesosphere above 60 km to offer a reliable statistics. Inclusion of GW effects leads, generally, to
colder simulated temperatures, which provide favorable conditions for cloud formation. This cooling in
the mesosphere is mainly produced via the dynamical channel due to GW-induced changes in the winds that
affect temperature through the thermal wind relation, rather than via the thermal channel due to direct
heating/cooling by dissipating GW harmonics. The latter clearly transpires in experiments with thermal
effects of the parameterized waves turned on and off (not shown). The direct thermal effects of GWs
increasingly grow with height and become important near the mesopause and above.

The vast majority of studies report on cloud observations in the Martian mesosphere below $\sim$80
km. \citet{SeftonNash_etal13} analyzed data from Mars Climate Sounder (MCS) on board the Mars Reconnaissance
Orbiter \citep{Graf_etal05, ZurekSmrekar07} (MRO) during dayside and nightside local times over two Martian years
and provided a global picture of high-altitude clouds. They found out that the distribution of clouds over latitude
and season does not appear to vary between each Martian year and that clouds occurred more often in low latitudes
during the aphelion season and concentrated around two midlatitude bands during perihelion. Using two different
observational modes, \citet{SeftonNash_etal13} showed that the latitudinal distributions of clouds varied little
between the different local times in the second half of the year. It must be noted that \citet{SeftonNash_etal13}
could not discriminate between CO$_2$ and water clouds. The majority of positively identified CO$_2$ cloud
observations took place in the first half of the year with only a few detections in the second half. On the other
hand, water ice clouds usually do not extend higher than $\sim$40 km except during perihelion, when they rise to
60-65 km. Therefore, all clouds observed above 70 km are likely not water ice clouds. The question regarding the
nature of these detected clouds is still open.  Our simulations illustrate that favorable conditions for CO$_2$
condenstation in the mesosphere exist in low latitudes throughout the year (Figure~\ref{fig:mean_gw}g). This is
because the mean temperature is lowest around the equator at all seasons
(Figure~\ref{fig:mean_fields_sol}a), while GW-induced temperature fluctuations are contrary small
(Figure~\ref{fig:mean_gw}a). The model also predicts a higher probability of cloud formation in midlatitudes of the
summer hemisphere during both solstices.
 
It is observationally challenging to determine the precise altitude of CO$_2$ ice clouds and there are
intrinsic limitations of the retrieval algorithms associated with CO$_2$ cross-sections
\citep{Maattanen_etal13}. Nevertheless, our modeling results can qualitatively be compared with
\citet{SeftonNash_etal13}'s analysis of the seasonal variation of Martian high-altitude clouds.  The
observations show that during northern hemisphere summer, clouds formed in the mesosphere more rarely
than during perihelion and located mainly around the equator. Previous analysis of the data from the
Thermal Emission Spectrometer (TES) on board the Mars Global Surveyor (MGS) \citep{Clancy_etal07} also
indicated that cloud occurrences were confined to a narrow latitude sector of $\pm 15^\circ$ during the
aphelion season ($L_s=30^\circ-150^\circ$). In agreement with observations, our simulations show higher
probabilities of cloud formation in low latitudes throughout all seasons (Figure~\ref{fig:mean_gw}g). The
latter is simply a consequence of the temperature minimum near the equatorial mesopause. The model
reproduces more favorable conditions for CO$_2$ condensation in the midlatitude regions during
wintertime. It agrees with observations in that mesospheric clouds occur more frequently during
perihelion (Figure~\ref{fig:mean_gw}g). Cold pockets with supersaturated temperatures occur in the model
only in less than 10\% of time at $\sim$80 km, but their probabilities grow with height. At 100 km,
maxima of $P$ of up to $\sim$18\% are seen in midlatitudes of the summer hemispheres
(Figure~\ref{fig:mean_gw}h), while higher up at $\sim$120 km these maxima exceed $\sim$30\% and
shift poleward (Figure~\ref{fig:mean_gw}i). Such behavior is due to growing with height amplitudes of GWs
and the associated temperature fluctuations (Figure~\ref{fig:mean_gw}a-c). The shift of maxima of cloud
probabilities first to middle and then to high latitudes is caused by the cold anomaly of the mean
temperature, which is induced by GWs in the mesosphere. The similar GW-induced cold summer mesopause
anomaly is well known in the atmosphere of Earth \citep{GarciaSolomon85}.

There are certain disagreements between the modeled cloud formation probabilities and existing
observations at high altitudes (80 km and above). In particular, the day-side observations of
\citet{SeftonNash_etal13} demonstrate a more symmetric with respect to the equator distribution of CO$_2$
clouds during the first part of the year. They do not show a ``pause" near the equinox around
$L_s=180^\circ$, which is clearly seen in Figures~\ref{fig:mean_gw}h,f. It is worth noting that the
superposition of the simulated patterns at 80 and 100 km is close to the superposition of the observed
night- and day patterns \citep[][Fig. 6]{SeftonNash_etal13} attributed to the 80 km altitude. Finally,
there is no statistically significant observational support for the predicted cloud formation probability
above 80 km. We, therefore, discuss possible reasons and shortcomings of the modeling methodology.

One source of uncertainty in our simulations is the assumed degree of supersaturation, which is currently
$35\%$ based on previous experimental constraints \citep{Glandorf_etal02}. However, a variable with
height supersaturation threshold is possible, which could modulate $P$ in our numerical experiments. This
variable threshold may reflect the microphysics of cloud formation, which implies an existence of nuclei
and strong dependence on their sizes. It is likely that the existence of cold pockets (the necessary
condition for cloud formation) is far from sufficient for clouds to form, especially in the upper
atmosphere. Thus, the lack of nuclei in in the upper atmosphere may prevent cloud formation.
If formed, ice particles must be small (being of submicron size) and clouds are too thin to be previously detected. 

%Discussion of radiative feedback effects and dust effects that we have not included.
In all our simulations, only probabilities of GW-induced clouds were calculated and, thus, no radiative
effect of such clouds were taken into account. Such radiative feedback has been considered, for example,
in the work by \citet{SiskindStevens06} in the Earth context. However, one can expect that, given that IR
CO$_2$ and GW thermal effects together dominate the energy budget of the mesosphere
\citep[e.g.,][]{Medvedev_etal15}, secondary radiative processes are likely to play a relatively minor
role in the cloud formation
%For example, the annual global mean cooling due to the combined effects of CO$_2$ and GWs is 
%$\sim 70-80$ K~sol$^{-1}$ above 80 km (Figure~\ref{fig:global_mean}). 
by producing local modulations of temperature.  A more comprehensive examination of the radiative feedback
processes in the Martian environment would require a two-way coupling between microphysics and small- and
large-scale dynamics. Interestingly, using a one-dimensional radiative-convective model, \citet{Mischna_etal00}
demonstrated that the lower atmospheric CO$_2$ clouds have a potential to produce an additional cooling of the
Martian surface by reflecting the incoming solar radiation. A further source of uncertainty is the use of
  one-dimensional atomic oxygen profile, which may affect neutral temperatures.

An obvious candidate for explaining mismatches between the modeling and observations is the specification of
sources in the GW parameterization. In the simulations, we assumed a globally uniform and constant with time
distribution of GW momentum fluxes. The magnitudes of the fluxes were chosen from observations to capture the
``background" effect of small-scale waves, as described in detail in our earlier works
\citep{Medvedev_etal11a,Medvedev_etal11b}.  Recently, using a high-resolution Martian GCM,
\citet{Kuroda_etal15,Kuroda_etal16} have shown a strong seasonal and latitudinal variation of GW momentum fluxes in
the lower atmosphere and, as a result, significant variations of GW-induced activity in the middle atmosphere.
Constraining wave sources is a logical next step in model development, which can potentially improve simulations of
clouds.
 
Finally, other limitations in the model can result in imperfections with the simulated mean fields and, as a
consequence, with erroneous estimates of cloud formation probability $P$. The Reviewer suggested that accounting
for radiative effects of water clouds and for more realistic dust scenario (mainly associated with its vertical
distribution) may affect the simulated $P$. These and other undertaken paths of MGCM sophistication, like
self-consistent modeling of water and aerosol cycles, can potentially bring observations and simualations of CO$_2$
clouds closer.

%\conclusions[Summary and Conclusions]  
\section{Summary and Conclusions}  
\label{sec:conclusion}

We presented simulations with the Max Planck Institute Martian General Circulation Model (MPI-MGCM)
\citep{Medvedev_etal13}, incorporating a whole atmosphere subgrid-scale gravity wave (GW)
parameterization of \citet{Yigit_etal08}, of distributions of mean fields, GW effects, and cloud
formation probabilities over one Martian year, assuming multi-year averaged observed dust distribution
with major dust storms removed. Model results are compared to a run without subgrid-scale effect
included.

Inclusion of effects of small-scale GWs 
facilitates CO$_2$ cloud formation in two ways. First, they cool down the upper atmosphere 
globally and, second, they create excursions of temperature well below the CO$_2$ condensation threshold 
in some parts of the middle atmosphere.  The main findings of this study are as follows.
\begin{enumerate}
\item GWs lead to $\sim$9\% colder global annual mean temperatures and even stronger temperature drops
  locally. Global annual mean GW-induced cooling of $-30$ K sol$^{-1}$ is comparable with that of due to
  radiative transfer by CO$_2$ molecules around 100 km and exceeds it above, reaching $-80$ K sol$^{-1}$
  around 140 km. GW-induced effects modulate the CO$_2$ cooling via changes in the background
  temperature.
\item Simulations reveal strong seasonal variations of GW effects in the upper mesosphere and lower
  thermosphere with solsticial maxima: Eastward GW drag peaks during the summer solstices and westward GW
  drag maximizes around the winter solstices with up to $\pm 1000$ m s$^{-1}$ sol$^{-1}$.  
\item Around the mesopause, GW-induced temperature fluctuations $|T^\prime|$ can exceed 20 K and
  the ice cloud formation probability ($P$) can be greater than $20\%$ locally.
\item Overall, GW temperature fluctuations substantially correlate with the cloud formation
  probability, in particular at middle- and high-latitudes in the upper mesosphere and
  mesopause region during all seasons.
\item Cloud formation exhibits strong seasonal variations larger than $30\%$, with summer solsticial maxima at
  high-latitudes in the mesosphere and around the mesospause.
\item The simulated seasonal variations of clouds probabilities in the mesosphere are in a reasonable
  agreement with previous detections of two distinct mesospheric types of clouds, i.e., equatorial
  and midlatitudes clouds. 
\end{enumerate} 

This study has shown that accounting for GW-induced temperature fluctuations in the Martian GCM
reproduces supersaturated cold temperatures in the upper mesosphere throughout all
seasons. GWs maintain globally cooler air, which is necessary for ice cloud formation, and help to
explain some features of the observed seasonal behavior of high-altitude CO$_2$ ice clouds.  Owing to
GW-induced globally colder temperatures and local temperature fluctuations, high-altitude clouds can form
from the upper mesosphere to the mesopause region, and occasionally even slightly above the mesopause. We
conclude that GW dynamical and thermal effects not only maintain the colder Martian mesosphere and lower
thermosphere, but also significantly contribute to the specific features of the observed high-altitude
clouds and their seasonal variations.

This study also puts forward new questions. Are our results concerning shaping the
seasonal behavior of ice clouds model-specific? Can CO$_2$ clouds form at altitudes above the mesopause,
as the simulations predict? How does the microphysics of cloud formation modify these predictions?
Further systematic modeling and obervational efforts have to be performed in order to address these open
questions.

\acknowledgements{
%\begin{acknowledgements}
 The modeling data supporting the figures presented in this paper can be obtained from EY
  (eyigit@gmu.edu). This work was partially supported by German Science Foundation (DFG) Grant
  HA3261/8-1. EY was funded by
  the National Science Foundation (NSF) grant AGS 1452137.
%\end{acknowledgements}
 }  

\begin{appendix}

\section{Martian parameters and seasons}
\label{sec:Martian}
%%% TABLES
%%%

Mars demonstrates in terms of planetary parameters some similarities as well as differences to Earth as summarized
in Table \ref{tb:1}.

\begin{table}[h]
\caption{Some key planetary parameters of Earth and Mars.}
\label{tb:1}
\begin{center}
\begin{tabular}{l|cc}
\hline
Planetary parameters     & Mars  & Earth \\
\hline
Mean solar distance [AU]     & 1.52   &  1    \\
Radius     [km]         & 3389  & 6370  \\
Length of a day [h]     & 24.65 &  24\\
Length of year [days]   & 687    & 365.5\\
Axial tilt [degrees]    & $25.19^\circ$ & $23.5^\circ $\\
Gravity    [m s$^{-2}$]  & 3.72   & 9.81\\
Eccentricity            & 0.0934   & 0.0167 \\
\hline
\end{tabular}
\end{center}
 % \begin{tablenotes}
 %      \small
 %    \item The solar distance designates the average distance from Sun, given in terms of AU $\sim 150$ million
 %      km. The axial tilt, or obliquity of the orbit, is measured with respect to the orbital plane. One solar day
 %      on Mars is referred to as one sol.
 %    \end{tablenotes}
% \belowtable{The solar distance designates the average distance from Sun, given in terms of AU $\sim 150$ million km. The axial tilt, or obliquity of the orbit, is measured with respect to the orbital plane. One solar day on Mars is referred to as one sol.} % Table Footnotes
%\multicolumn{2}{|p{3cm}|}
\end{table}

In planetary atmospheres, one ``sol" refers to the duration of a solar day on Mars. The length of a day is
  longer on Mars than on Earth. One Martian sol is about 24 hours and 39 minutes (i.e., 24.65h), thus slightly
  longer than and an Earth day. One Martian year is 687 days long or 669 Martian sols. Due to different
  eccentricities of Mars and Earth, their distance can vary significantly over the course of their orbital motion
  around Sun. Martian seasons are described by the solar longitude $L_s$. In our modeling, by convention, $L_s=0$
vernal equinox, $L_s=90^\circ$ is northern hemisphere solstice (aphelion), $L_s=180^\circ$ is autumnal equinox, and
$L_s=270^\circ$ is northern hemisphere winter solstice (perihelion).

\end{appendix}

%
% FIGURES 
%

\end{document}